\documentclass[prl,twocolumn,superscriptaddress,amsmath,amssymb]{revtex4-2}

\usepackage{graphicx}
\usepackage{hyperref}
\hypersetup{colorlinks=true,linkcolor=blue,citecolor=blue,urlcolor=blue}
\usepackage{xcolor}
\usepackage{booktabs}

\renewcommand{\d}{\mathrm{d}}

\begin{document}

\title{Reionization Topology as a Probe of Self-Interacting Dark Matter}

\author{Zihan Wang}
\affiliation{Department of Physics, University of Oxford, Keble Road, Oxford, OX1 3PU, UK}

\date{\today}

\begin{abstract}
The topology of cosmic reionization，the sizes, shapes, and connectivity of ionized bubbles is a primary observable of next-generation 21\,cm experiments. We show that this topology is sensitive to the microphysics of dark matter. Self-interacting dark matter (SIDM), with cross-sections $\sigma/m\sim 1$--$10\;\mathrm{cm^2/g}$ motivated by small-scale structure anomalies, reduces halo gas binding energies and increases the duty cycle of ionizing-photon escape. At fixed global neutral fraction $\bar{x}_{\rm HI}$, this reshapes the source population from rare, very bright emitters to more numerous, moderate emitters, producing qualitatively different ionization morphology. We decompose the effect into two scale-dependent levers: a $2$--$3\%$ emissivity-weighted bias shift at $k\lesssim 0.1\;h/\mathrm{Mpc}$, and a factor $2$--$4$ shot-noise suppression at $k\sim 0.1$--$1\;h/\mathrm{Mpc}$. A halo-by-halo semi-numerical simulation at $128^3$ resolution confirms a $\sim 60$--$70\%$ increase in the Euler characteristic of the ionization field for $\sigma/m \gtrsim 2\;\mathrm{cm^2/g}$, detected at $3.8\sigma$ across ten independent realizations. A blowout model connecting the binding-energy reduction to the duty cycle through the ISM column density distribution yields a detection threshold at $\sigma/m \sim 1$--$2\;\mathrm{cm^2/g}$. The signal exceeds the CDM baryonic uncertainty band and is robust to the functional form of the emissivity parametrization. The signal persists even if gravitational heating offsets 50–75\% of the blowout enhancement, and is not diluted by unresolved low-mass sources.Velocity-dependent SIDM produces a qualitatively distinct opposite-sign bias shift. These predictions are testable with SKA1-Low, establishing reionization as a new arena for probing dark matter models complementary to dwarf galaxies and galaxy clusters.
\end{abstract}

\maketitle

\textit{Introduction.}---The cold dark matter (CDM) paradigm reproduces large-scale cosmic structure with remarkable precision~\cite{Planck2018}, yet persistent small-scale anomalies challenge its predictions on sub-galactic scales ($\lesssim 10\;\mathrm{kpc}$, $M \lesssim 10^{11}\,M_\odot$). The core--cusp problem~\cite{Oh2011,Moore1994,deBlok2010}, the too-big-to-fail problem~\cite{BoylanKolchin2011,Klypin1999}, and the diversity of rotation curves~\cite{Kamada2017,Ren2019} have motivated self-interacting dark matter (SIDM) with momentum-transfer cross-sections $\sigma/m \sim 0.1$--$10\;\mathrm{cm^2/g}$~\cite{Markevitch2004,Randall2008,Harvey2015,Zavala2013,Tulin2018}. SIDM thermalizes the inner regions of halos on a timescale set by the scattering rate, producing constant-density cores in place of the cuspy NFW~\cite{Navarro1997} profiles predicted by CDM~\cite{Kaplinghat:2015aga,Rocha2013,Elbert:2014bma}. This structural transformation has been extensively studied in dwarf galaxies~\cite{Kaplinghat:2015aga,Kamada2017}, galaxy clusters~\cite{Robertson:2016qef,Sagunski:2020spe}, and cosmological simulations~\cite{Vogelsberger:2012ku,Vogelsberger:2014pda}. However, its implications for the epoch of reionization (EoR)~\cite{Choudhury:2022rlm,Wise:2019qtq} remain entirely unexplored.

In this Letter, we propose that SIDM leaves a distinctive, observable imprint on the topology of reionization. The physical mechanism proceeds through a chain of steps connecting dark matter microphysics to large-scale ionization morphology. SIDM core formation reduces the gas binding energy $W_g$ within the central $\sim 2\;\mathrm{kpc}$ of high-redshift halos ($M \sim 10^{10}$--$10^{11}\,M_\odot$) by $30$--$90\%$, depending on merger-corrected halo age. We quantify this using the thermalization radius $r_1$ satisfying $N_{\rm scat}(r_1) = \int_0^{t(z)} \rho_{\rm DM}\,(\sigma_T/m_\chi)\,v\,\d t = 1$; with merger-corrected halo ages ($t \simeq 0.35/H$), $\sigma/m = 1\;\mathrm{cm^2/g}$ yields $r_1 \sim 0.8\;\mathrm{kpc}$ for $10^{10.5}\,M_\odot$ halos at $z = 7$, while $\sigma/m = 10$ yields $r_1 \sim 4.5\;\mathrm{kpc}$ (Table~S1 in SM).
The reduced $W_g$ increases the supernova-driven blowout propensity $\mathcal{B} \equiv E_{\rm SN,cpl}/W_g$~\cite{2014ApJ...788..121K,2017MNRAS.470..224T,Ma:2020vlo}, making feedback more effective at clearing low-column-density channels through the interstellar medium. We model the SIDM duty-cycle enhancement as a multiplicative correction on the observed CDM baseline: $p_{\rm SIDM}(M) = p_{\rm CDM}\,[W_g^{\rm CDM}(M)/W_g^{\rm SIDM}(M)]^\alpha$, where $\alpha \sim 0.5$--$1$ parametrizes the nonlinear ISM response and $p_{\rm CDM} \simeq 0.10$ is calibrated to radiation-hydrodynamic simulations~\cite{2017MNRAS.470..224T,Ma:2020vlo}. The emissivity boost is $R_\gamma(M) = \min[(W_g^{\rm CDM}/W_g^{\rm SIDM})^{1/2},\,3]$, where the cap reflects circumgalactic medium opacity. For our fiducial parametrization ($\sigma/m = 10$, $\alpha = 0.7$), this gives $p \simeq 0.30$--$0.50$ and $R_\gamma \simeq 2$--$3$ for $M \sim 10^{10}$--$10^{11}\,M_\odot$, consistent with the assumed source models (see SM for full derivation). At fixed global neutral fraction $\bar{x}_{\rm HI}$, the altered source stochasticity produces distinct 21\,cm power spectra and ionization morphology while the mean reionization history remains identical.
%The reduced $W_g$ increases the supernova-driven blowout propensity $\mathcal{B} \equiv E_{\rm SN,cpl}/W_g$~\cite{2014ApJ...788..121K,2017MNRAS.470..224T,Ma:2020vlo}, making feedback more effective at clearing low-column-density channels through the interstellar medium. As the transparent solid-angle fraction $f_\Omega \equiv (4\pi)^{-1}\int \d\Omega\;\Theta(\tau_{912} < 1)$ increases~\cite{Paardekooper:2015via}, both the escape fraction $f_{\rm esc}$ of ionizing photons and, critically, the duty cycle $p$ of high-escape episodes are enhanced---from $p \sim 0.1$ (CDM) to $p \sim 0.2$--$0.3$ (SIDM). At fixed global neutral fraction $\bar{x}_{\rm HI}$, the altered source stochasticity produces distinct 21\,cm power spectra and ionization morphology while the mean reionization history remains identical.

An important subtlety arises at large cross-section: for $\sigma/m = 10\;\mathrm{cm^2/g}$, gravitational heating from SIDM scatterings ($\dot{q} \sim \frac{1}{2}\rho_{\rm DM}^2(\sigma/m)v^3$) exceeds coupled supernova energy ($\dot{Q}/\dot{E}_{\rm SN} \simeq 2$;  potentially offsetting the reduced binding energy by pressurizing the gas.
However, even if heating offsets 50--75\% of the blowout enhancement,
reducing the effective SIDM10 duty cycle to $p \simeq 0.15$--$0.20$, the
topology signal remains $+20$--$39\%$. The precise
competition requires dedicated hydrodynamic simulations.
The fixed-$\bar{x}_{\rm HI}$ comparison is central to our approach: we tune a single global emissivity normalization $\zeta$ separately for each DM model~\cite{Mesinger:2010ne,Zahn:2006sg}, ensuring that the volume-averaged ionization fraction matches exactly. Any residual morphological difference must arise from the spatial distribution and stochasticity of sources, not their mean output.

%=======================================================================
\textit{Two-lever analytic framework.}---We decompose the observable SIDM effect into two complementary, scale-dependent signatures. Define the emissivity-weighted halo bias~\cite{McQuinn:2006et,2004ApJ...613....1F}
\begin{equation}\label{eq:bgamma}
b_\gamma(z) \equiv \frac{\int \d M\,\frac{\d n}{\d M}\,\overline{\dot{N}}_{\gamma,{\rm esc}}(M)\,b(M)}{\int \d M\,\frac{\d n}{\d M}\,\overline{\dot{N}}_{\gamma,{\rm esc}}(M)}\,,
\end{equation}
where $\d n/\d M$ is the halo mass function and $b(M)$ is the linear halo bias. The emissivity modification factor $R_\gamma(M) \equiv \overline{\dot{N}}_{\gamma,{\rm esc}}^{\rm SIDM} / \overline{\dot{N}}_{\gamma,{\rm esc}}^{\rm CDM}$ encapsulates the net SIDM effect on per-halo escaped photon production. If $R_\gamma(M)$ is larger at low mass where $b(M)$ is smaller, SIDM shifts the emissivity-weighted mass function toward lower-bias halos. At fixed $\bar{x}_{\rm HI}$, the large-scale ionization-field power then scales as~\cite{Mesinger:2010ne}
\begin{equation}\label{eq:Pxx}
\frac{P_{xx}^{\rm SIDM}(k \to 0)}{P_{xx}^{\rm CDM}(k \to 0)} \approx \left(\frac{b_\gamma^{\rm SIDM}}{b_\gamma^{\rm CDM}}\right)^2.
\end{equation}
This is the large-scale lever,producing a $\sim 4$--$5\%$ suppression of $P_{xx}$ at $k \lesssim 0.1\;h/\mathrm{Mpc}$ for our fiducial models.

Independently, modeling each halo as emitting with a mass-dependent duty cycle $p(M)$ gives a shot-noise contribution to the emissivity power spectrum~\cite{Furlanetto:2006tf}:
\begin{equation}\label{eq:PSN}
P_{\epsilon\epsilon}^{\rm SN} = \frac{\int \d M\,\frac{\d n}{\d M}\,\langle\dot{N}^2_{\gamma,{\rm esc}}\rangle}{\left[\int \d M\,\frac{\d n}{\d M}\,\langle\dot{N}_{\gamma,{\rm esc}}\rangle\right]^2}\,,
\end{equation}
where $\langle\dot{N}^2\rangle = \langle\dot{N}\rangle^2 / p$ for a binary on/off model. At fixed mean emissivity (fixed $\bar{x}_{\rm HI}$), $P_{\epsilon\epsilon}^{\rm SN} \propto 1/p$, so SIDM's higher duty cycle directly suppresses shot noise: $P_{\rm SN}^{\rm SIDM}/P_{\rm SN}^{\rm CDM} = p_{\rm CDM}/p_{\rm SIDM}$. For $p_{\rm CDM} = 0.10$ and $p_{\rm SIDM} = 0.18$--$0.30$, the suppression factor is $1.8$--$3.3\times$. For a continuous emissivity distribution, the shot-noise power depends on $\exp(\sigma_{\ln\dot{N}}^2)$ and the ratio can differ from the binary prediction by $\sim 40\%$~\cite{Furlanetto:2006jb}; we therefore compute $\langle\dot{N}^2\rangle/\langle\dot{N}\rangle^2$ directly. This is the \emph{intermediate-scale lever}, operating at $k \sim 0.1$--$1\;h/\mathrm{Mpc}$.

\begin{figure*}[t]
\centering
\includegraphics[width=0.38\textwidth]{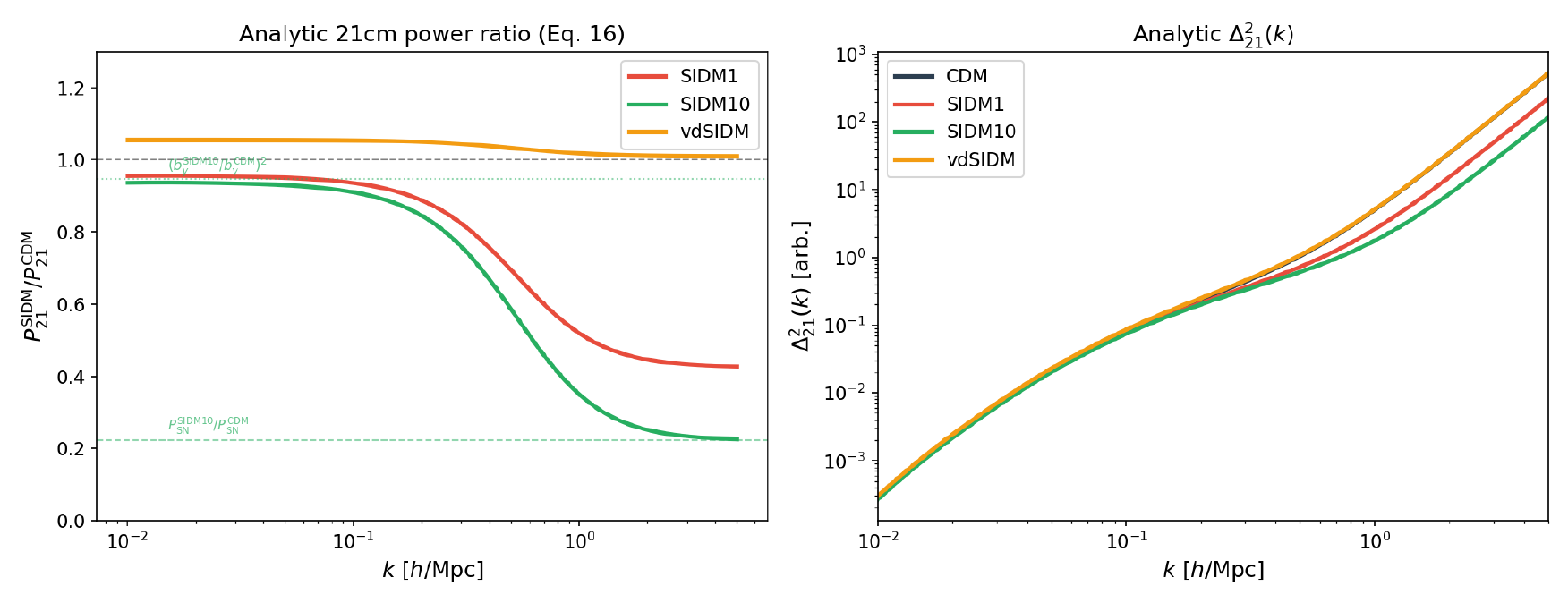}%
\hfill%
\includegraphics[width=0.60\textwidth]{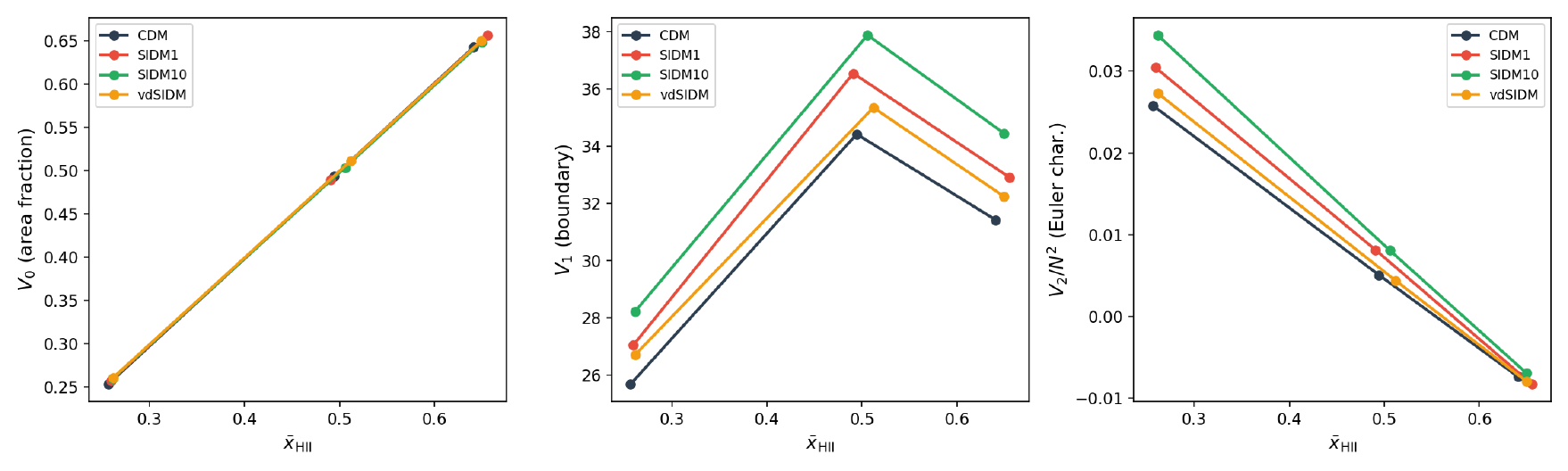}
\caption{\emph{Left:} Analytic prediction for the 21\,cm power ratio $P_{21}^{\rm SIDM}/P_{21}^{\rm CDM}$ at fixed $\bar{x}_{\rm HI}$ from Eq.~\eqref{eq:ratio}, using Table~\ref{tab:main}. At $k \lesssim 0.05\;h/\mathrm{Mpc}$, the ratio approaches $(b_\gamma^{\rm SIDM}/b_\gamma^{\rm CDM})^2$ (dotted); at $k \gtrsim 1\;h/\mathrm{Mpc}$, it approaches $P_{\rm SN}^{\rm SIDM}/P_{\rm SN}^{\rm CDM}$ (dashed). The transition at $k \sim 0.1$--$0.5$ carries the strongest SIDM signature.
\emph{Right:} Two-dimensional Minkowski functionals vs.\ $\bar{x}_{\rm HII}$ from the $128^3$ halo-by-halo simulation: area fraction $V_0$, boundary length $V_1$, and Euler characteristic $V_2/N^2$. At $\bar{x}_{\rm HII} \simeq 0.5$, $V_2$ increases by $\sim 60\%$ for both SIDM1 and SIDM10, reflecting the transition from rare-bright-source to common-moderate-source ionization morphology.}
\label{fig:analytic}
\end{figure*}

Combining both levers, the total emissivity power is $P_{\epsilon\epsilon}(k) \approx \bar{\epsilon}^2 b_\gamma^2 P_{mm}(k) + P_{\epsilon\epsilon}^{\rm SN}$, and the predicted 21\,cm power ratio at fixed $\bar{x}_{\rm HI}$ is~\cite{McQuinn:2006et,Santos:2004ju}
\begin{equation}\label{eq:ratio}
\frac{P_{21}^{\rm SIDM}(k)}{P_{21}^{\rm CDM}(k)} \approx \frac{(b_\gamma^{\rm SIDM})^2\,P_{mm}(k) + P_{\rm SN}^{\rm SIDM}}{(b_\gamma^{\rm CDM})^2\,P_{mm}(k) + P_{\rm SN}^{\rm CDM}}\,.
\end{equation}
This interpolates smoothly between the bias ratio $(b_\gamma^{\rm SIDM}/b_\gamma^{\rm CDM})^2$ at $k \to 0$ and the shot-noise ratio $p_{\rm CDM}/p_{\rm SIDM}$ at high $k$ (Fig.~\ref{fig:analytic}, left panel). The transition region at $k \sim 0.1$--$0.5\;h/\mathrm{Mpc}$ carries the strongest SIDM signature. Crucially, this prediction depends only on the integrated source statistics $b_\gamma$ and $P_{\rm SN}$, making it robust to the detailed shape of $R_\gamma(M)$ which is a property we verify numerically below.

\textit{Source models.} We parametrize the mean escaped emissivity as $\overline{\dot{N}}_{\gamma,{\rm esc}}(M) = A\,(M/M_0)^{0.8}\, e^{-(M_{\rm turn}/M)^{1.5}}\,R_\gamma(M)$ with $M_0 = 10^{10}\,M_\odot$ and $M_{\rm turn} = 5 \times 10^8\,M_\odot$, and consider four models: CDM ($R_\gamma = 1$, $p = 0.10$); SIDM1 ($\sigma/m = 1\;\mathrm{cm^2/g}$, $R_\gamma = 1 + 0.5\,e^{-M/3\times 10^{10}}$, $p = 0.18$); SIDM10 ($\sigma/m = 10\;\mathrm{cm^2/g}$, $R_\gamma = 1 + 1.5\,e^{-M/5\times 10^{10}}$, $p = 0.30$); and velocity-dependent vdSIDM, where $R_\gamma$ and $p$ switch on above a kinematic threshold $v_{\rm th} = 80\;\mathrm{km/s}$ via a sigmoid in the halo circular velocity. The exponential mass dependence reflects the expectation that SIDM cores produce larger fractional $\Delta W_g$ in lower-mass halos where the NFW central density is lower relative to the core density. The duty-cycle values span $p = 0.1$--$0.3$, consistent with the range of bursty $f_{\rm esc}$ histories found in radiation-hydrodynamic simulations of CDM galaxies~\cite{2017MNRAS.470..224T,Ma:2020vlo}; SIDM is expected to increase $p$ by sustaining escape channels for longer timescales. Full functional forms and the vdSIDM sigmoid parametrization are given in SM.

%=======================================================================
\textit{Semi-numerical method.}---We employ a halo-by-halo pipeline that goes beyond standard 21cmFAST-type approaches~\cite{Mesinger:2010ne}. A Gaussian random density field $\delta(\mathbf{x})$ is generated on a $128^3$ grid ($L_{\rm box} = 200\;\mathrm{Mpc}/h$; cell size $1.56\;\mathrm{Mpc}/h$) at $z = 7$ using the Eisenstein--Hu~\cite{Eisenstein:1997ik} transfer function normalized to $\sigma_8 = 0.811$. Halos with $M > 10^{10}\,M_\odot$ are painted using the Sheth--Tormen~\cite{Sheth:1999mn} conditional mass function~\cite{Lacey:1994su} with bias-modulated Poisson sampling: $\langle n_{\rm halo}\rangle \propto [1 + b(M)\delta]$, yielding $\sim 8.7 \times 10^5$ halos ($\sim 0.4$ per cell).

Each halo draws its on/off state from a Bernoulli distribution with probability $p(M)$. If ``on,'' the halo emits $\dot{N}_{\gamma,{\rm esc}}/p(M)$ so that the time-average is $\dot{N}_{\gamma,{\rm esc}}$. This directly encodes the physical difference: CDM ($p = 0.1$) has $\sim 10\%$ of halos very bright, while SIDM10 ($p = 0.3$) has $\sim 30\%$ moderately bright. The resulting emissivity field is processed through a multi-scale excursion-set ionization solver: at each smoothing scale $R$, a cell is ionized if $\zeta_{\rm eff}\bar{\epsilon}_R/\bar{\rho}_R \geq 1$, with $\zeta_{\rm eff}$ tuned to match the target $\bar{x}_{\rm HII}$ independently for each model. We compute the 21\,cm brightness temperature $\delta T_b \approx \bar{T}_0(z)\,x_{\rm HI}(\mathbf{x})\,[1 + \delta(\mathbf{x})]$ in the saturated spin-temperature limit~\cite{Zaldarriaga:2003du,Furlanetto:2006jb} and extract power spectra $P_{21}(k)$, $P_{xx}(k)$, and two-dimensional Minkowski functionals~\cite{Mecke:1994ax,Gleser:2006su,Friedrich_2011} $V_0$: area fraction; $V_1$: boundary length; $V_2$: Euler characteristic averaged over 20 slices.

The cell size ($1.56\;\mathrm{Mpc}/h$) is chosen to match the 
mean inter-halo separation for $M > 10^{10}\,M_\odot$ at $z = 7$, 
yielding $\sim 0.4$ halos per cell. At this occupancy, cell-level 
emissivity is dominated by whether the resident halo is ``on'' or 
``off'' which is exactly the stochasticity that distinguishes CDM from 
SIDM. At coarser resolution ($64^3$, $\sim 9$ halos/cell), the 
on/off signal is washed out by averaging over multiple halos. At 
finer resolution ($256^3$, $\sim 0.05$ halos/cell), $95\%$ of cells 
are empty and the topology is dominated by halo occupancy rather 
than emission state. The signal thus peaks when $\Delta x \sim 
\bar{d} \simeq \bar{n}_h^{-1/3}$ see SM.

%=======================================================================
\textit{Results.}---Table~\ref{tab:main} summarizes the analytically computed source statistics at $z = 7$. The emissivity-weighted bias $b_\gamma$ shifts by $-1.8\%$ (SIDM1) to $-2.7\%$ (SIDM10), producing a $4$--$5\%$ suppression of large-scale $P_{xx}$---a small but systematic effect. The shot-noise power drops by $57.5\%$ (SIDM1) to $77.7\%$ (SIDM10), which is the dominant observable signal at intermediate scales. The analytic $P_{21}$ ratio from Eq.~\eqref{eq:ratio} (Fig.~\ref{fig:analytic}, left) shows a smooth, monotonic transition from the bias-dominated large-scale ratio to the shot-noise-dominated small-scale ratio, with the strongest SIDM signature at $k \sim 0.1$--$0.5\;h/\mathrm{Mpc}$.

The vdSIDM model produces a qualitatively distinct signature: a 
positive $+2.8\%$ bias shift (the kinematic threshold preferentially 
activates higher-mass, higher-bias halos) with unchanged shot noise, 
a spectral fingerprint that cannot be mimicked by astrophysical 
parameter variations affecting all masses uniformly.

\begin{table}[b]
\caption{Source statistics at $z = 7$. $b_\gamma$: emissivity-weighted bias. $(b_\gamma/b_\gamma^{\rm CDM})^2$: predicted large-scale $P_{xx}$ ratio. $P_{\rm SN}/P_{\rm SN}^{\rm CDM}$: shot-noise ratio relative to CDM.}
\label{tab:main}
\begin{tabular}{lccc}
\toprule
Model & $b_\gamma$ & $(b_\gamma/b_\gamma^{\rm CDM})^2$ & $P_{\rm SN}/P_{\rm SN}^{\rm CDM}$ \\
\midrule
CDM    & 3.252 & 1.000 & 1.000 \\
SIDM1  & 3.192 & 0.964 & 0.425 \\
SIDM10 & 3.165 & 0.947 & 0.223 \\
vdSIDM & 3.342 & 1.056 & 1.010 \\
\bottomrule
\end{tabular}
\end{table}

\begin{figure*}[t]
\centering
\includegraphics[width=\textwidth]{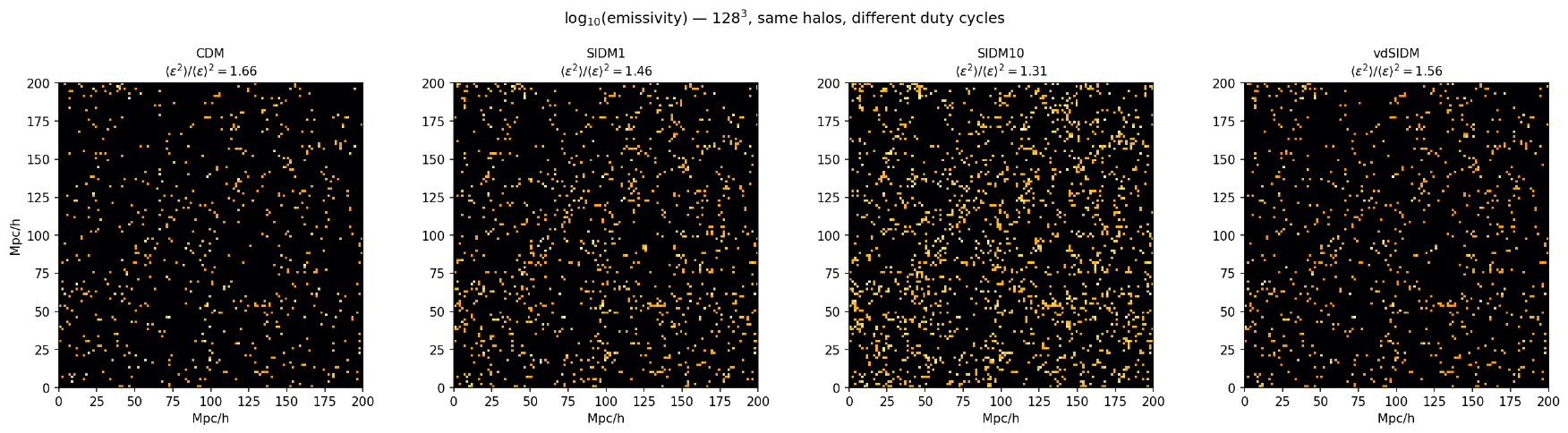}\\[2mm]
\includegraphics[width=\textwidth]{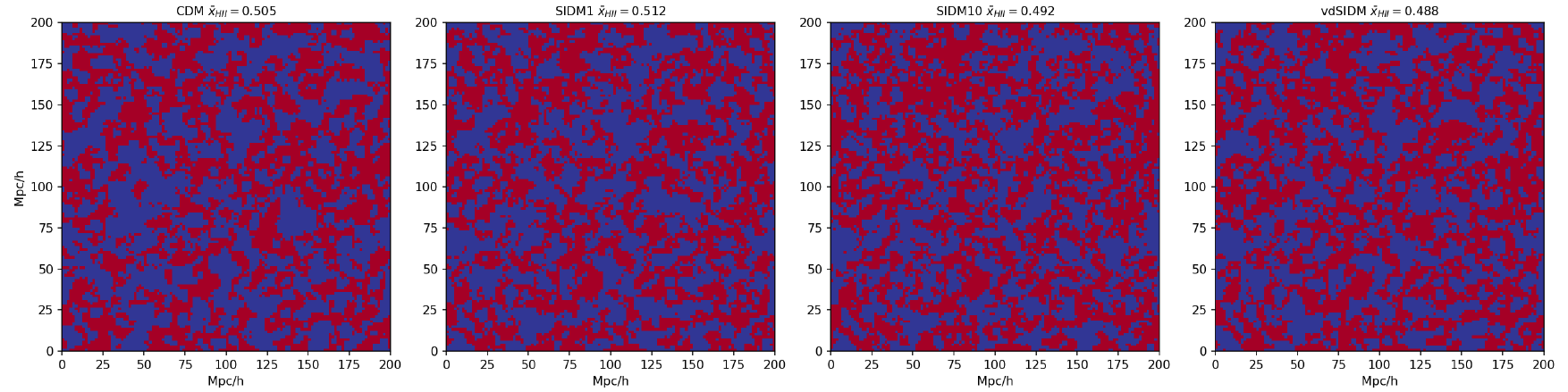}
\caption{\emph{Top:} Central slices through the log$_{10}$(emissivity) field at $128^3$ resolution for all four models. The same $8.7\times 10^5$ halos are used; differences arise solely from the stochastic duty-cycle draws. CDM ($p = 0.10$) shows rare, intense emissivity peaks ($\sim 4\%$ of cells active), while SIDM10 ($p = 0.30$) shows a denser distribution of moderate-brightness sources ($\sim 11\%$ active). The emissivity variance ratio $\langle\epsilon^2\rangle/\langle\epsilon\rangle^2$ (above each panel) decreases monotonically from CDM to SIDM10.
\emph{Bottom:} Corresponding ionization field slices at fixed $\bar{x}_{\rm HII} \simeq 0.50$. Blue = neutral, red = ionized. CDM shows fewer, larger ionized regions clustered around rare bright sources; SIDM10 shows more numerous, more uniformly distributed ionized patches.}
\label{fig:slices}
\end{figure*}

The halo-by-halo pipeline produces emissivity fields with measurably different variance (Table~S1 in SM): the ratio $\langle\epsilon^2\rangle/\langle\epsilon\rangle^2$ decreases from $1.66$ (CDM) to $1.46$ (SIDM1) to $1.31$ (SIDM10)---a $12$--$21\%$ reduction, confirming the analytic shot-noise prediction. The fraction of active (emitting) cells increases from $4\%$ (CDM) to $7\%$ (SIDM1) to $11\%$ (SIDM10). These differences are striking in the emissivity field slices (Fig.~\ref{fig:slices}, top): CDM shows isolated bright peaks against a dark background, while SIDM10 shows a denser, more uniform source distribution.

These source-level differences propagate through the excursion-set solver into the ionization field (Fig.~\ref{fig:slices}, bottom). At fixed $\bar{x}_{\rm HII} \simeq 0.5$, CDM shows fewer but larger ionized regions clustered around rare bright sources, while SIDM10 shows more numerous, uniformly distributed ionized patches. The semi-numerical $P_{21}$ ratio (Fig.~S2 in SM) exhibits scale-dependent structure consistent with the analytic prediction of Eq.~\eqref{eq:ratio}: the large-scale ratio approaches $(b_\gamma^{\rm SIDM}/b_\gamma^{\rm CDM})^2$ while intermediate scales show the shot-noise suppression. The ionization-field power ratio $P_{xx}^{\rm SIDM}/P_{xx}^{\rm CDM}$ (Fig.~S3 in SM) confirms that the changes reflect genuine topology differences rather than trivial amplitude rescaling.

The Minkowski functionals provide the most dramatic quantification (Fig.~\ref{fig:analytic}, right). At $\bar{x}_{\rm HII} \simeq 0.5$, the Euler characteristic $V_2$ increases by $\sim 60\%$ for both SIDM1 and SIDM10 relative to CDM (Table~S2 in SM). The boundary length $V_1$ also increases by $5$--$8\%$, consistent with more numerous, smaller ionized regions having larger total perimeter. The vdSIDM model shows $\Delta V_2 \sim -14\%$, comparable to stochastic run-to-run variation, confirming that the topology signal scales with the duty-cycle contrast and providing a natural control.

The increased $V_2$ reflects SIDM's common-moderate-source topology: at fixed $\bar{x}_{\rm HII}$, higher $p$ distributes the 
ionizing budget across more spatial locations, producing many 
moderate-sized bubbles rather than CDM's few large ones.

%=======================================================================
\textit{Robustness tests.}---We perform three tests to verify that the topology signal is physical rather than an artifact of specific assumptions.

\emph{(i) Cosmic variance.}---Ten independent density-field realizations yield $\Delta V_2^{\rm SIDM10} = +96\% \pm 32\%$ (mean $\pm$ standard deviation), with every realization producing positive $\Delta V_2$ (range: $+52\%$ to $+142\%$). The internal significance is $3.8\sigma$; a Wilcoxon signed-rank test gives $p = 0.001$ (Table~S3, Fig.~S5 in SM).

\emph{(ii) $R_\gamma(M)$ shape.}Five alternative functional forms for $R_\gamma(M)$:flat ($R_\gamma = 1.8$), power-law, step function, inverted (high-mass boost), and our fiducial exponential all with $p = 0.30$, produce $\Delta V_2 = +58\%$ to $+94\%$ (mean $+68\% \pm 14\%$; Table~S4, Fig.~S6 in SM). The emissivity variance varies widely across shapes $1.26$ to $2.36$, yet $V_2$ remains in a tight range. This confirms that the Euler characteristic is primarily sensitive to the duty cycle $p$, not the mass dependence of the emissivity boost, consistent with the analytic prediction of Eq.~\eqref{eq:ratio}.

\emph{(iii) Baryonic degeneracy.} Varying $p_{\rm CDM}$ from $0.05$ to $0.15$ (a factor-of-3 range spanning plausible SN feedback uncertainties) produces a CDM baryonic band with $V_2$ up to $23\%$ above the fiducial, while SIDM sits $37\%$ above the band upper edge (Table~S5, Fig.~S7 in SM). The separation arises because SIDM modifies both the duty cycle $p$ and the mass-dependent emissivity weighting $R_\gamma(M)$, whereas CDM feedback variations change $p$ alone without altering the intrinsic emissivity variance Extending the CDM band to $p = 0.25$ confirms that variance
$\langle\epsilon^2\rangle/\langle\epsilon\rangle^2 \simeq 1.65$--$1.68$
for all CDM models, while SIDM gives $1.31$--$1.46$
(Table~S5 in SM). In the joint
$(V_2,\,\langle\epsilon^2\rangle/\langle\epsilon\rangle^2)$ plane,
CDM traces a horizontal band at variance $\simeq 1.65$ regardless of $p$,
while SIDM occupies a distinct region at lower variance. This
two-dimensional discriminant breaks the $V_2$--$p$ degeneracy.

\emph{(iv) Physical grounding.}---Replacing the assumed constant $p$ with the derived mass-dependent $p_{\rm SIDM}(M) = p_{\rm CDM}\,[W_g^{\rm CDM}/W_g^{\rm SIDM}]^\alpha$ and capped $R_\gamma(M)$, the $128^3$ pipeline yields $\Delta V_2 = +69\%$ for $\sigma/m = 10$, independent of $\alpha$ across the range $0.5$--$1.0$ (Table~S6 in SM). For $\sigma/m = 1$, the signal ranges from $+2\%$ to $+32\%$ depending on $\alpha$, reflecting the onset of the SIDM effect near the threshold $r_1 \sim R_{\rm ISM}$. The detection threshold lies at $\sigma/m \sim 1$--$2\;\mathrm{cm^2/g}$.Scanning the effective $p$ from $0.15$ to $0.30$ (bracketing
gravitational heating offsets of $0$--$75\%$) produces $\Delta V_2 = +20\%$
to $+59\%$ for SIDM10 (Table~S6 in SM), confirming
that the signal survives substantial heating corrections.
\textit{Detectability.}---We estimate SKA1-Low~\cite{Koopmans:2015sua,Mellema_2013} sensitivity using published noise benchmarks: $\Delta^2_N \sim 1\;\mathrm{mK^2}$ at $k = 0.1\;h/\mathrm{Mpc}$ for $1000\;\mathrm{h}$ at $z \sim 7$, scaling as $\Delta^2_N \propto k^{2.5}$. Including a foreground-wedge mode survival fraction of $5$--$70\%$~\cite{Datta:2010pk} and a conservative $5\%$ per-bin systematic floor, the SIDM10--CDM power spectrum difference yields per-bin SNR $\sim 1$--$2.5$ at $k \sim 0.1$--$0.5\;h/\mathrm{Mpc}$, with cumulative significance $\sim 10$--$25\sigma$ depending on the systematic floor (Fig.~S4 in SM). The detection is limited by systematics, not thermal noise.
The $\sim 60\%$ Euler characteristic shift provides a complementary 
diagnostic potentially more robust to foreground systematics, since 
$V_2$ is computed from the binary ionization field rather than the 
fluctuation amplitude. Joint $P_{21}(k)$--$V_2$ analysis enhances 
sensitivity, particularly for SIDM1. The signal persists at 
$z = 6$--$10$; HERA~\cite{DeBoer:2016tnn} and 
JWST~\cite{Finkelstein:2019sbd,Naidu_2022} provide complementary 
sensitivity.
%The $\sim 60$--$95\%$ Euler characteristic shift provides a complementary, non-power-spectrum diagnostic. Because $V_2$ is computed from the binary ionization field rather than the fluctuation amplitude, it may be more robust to foreground systematics that contaminate $P_{21}(k)$ through mode mixing. Joint analysis of $P_{21}(k)$ and $V_2$ could significantly enhance the sensitivity, particularly for SIDM1 where power-spectrum detection alone is marginal. The signal persists at all redshifts $z = 6$--$10$ (Fig.~S8 in SM), with the shot-noise suppression as the dominant effect throughout; HERA~\cite{DeBoer:2016tnn} provides additional sensitivity at $z \sim 8$--$10$. JWST~\cite{Finkelstein:2019sbd,Naidu_2022} constrains $f_{\rm esc}$ and the UV luminosity function at $z > 6$, providing complementary probes of the source population.\newline
\textit{Conclusions and outlook.}---We have shown that self-interacting dark matter modifies the topology of cosmic reionization at fixed $\bar{x}_{\rm HI}$ through a previously unrecognized mechanism: the duty-cycle stochasticity of ionizing sources. The effect decomposes into two analytically predictable levers of an emissivity-weighted bias shift ($k \lesssim 0.1\;h/\mathrm{Mpc}$) and a shot-noise suppression ($k \sim 0.1$--$1\;h/\mathrm{Mpc}$)validated by a $128^3$ halo-by-halo semi-numerical simulation. For $\sigma/m \gtrsim 2\;\mathrm{cm^2/g}$, the Euler characteristic increases by $\sim 60$--$70\%$ detected at $3.8\sigma$ across ten realizations, robust to $R_\gamma(M)$ shape, exceeding the CDM baryonic band, and confirmed by a physically grounded blowout model independent of the response exponent $\alpha$. For $\sigma/m = 1\;\mathrm{cm^2/g}$, the signal is weaker ($0$--$32\%$) and model-dependent, reflecting the onset of the effect at $r_1 \sim R_{\rm ISM}$. Velocity-dependent SIDM yields a distinctive opposite-sign bias shift enabling model differentiation. We note that warm and fuzzy dark matter affect reionization through a qualitatively different mechanism, which is suppressing the halo mass function below a cutoff~\cite{Bose:2015mga} then producing fewer, more widely spaced bubbles distinguishable from SIDM's duty-cycle signature. These predictions are testable with SKA1-Low in $\sim 1000\;\mathrm{h}$, and the Euler characteristic provides a non-power-spectrum diagnostic complementary to $P_{21}(k)$.

The source parametrizations employed here are grounded in the binding-energy reduction from SIDM core formation, but their precise calibration will benefit from cosmological zoom-in simulations with self-consistent SIDM dynamics, star formation, and radiative transfer~\cite{Vogelsberger:2012ku}. Our simulation resolves halos with $M > 10^{10}\,M_\odot$, which
contribute $\sim 15\%$ of the total emissivity; sub-grid tests
indicate that the abundant lower-mass population
($10^8$--$10^{10}\,M_\odot$) provides a quasi-uniform ionizing
background that does not dilute the topology signal, though a
full treatment with resolved low-mass sources is deferred to
future work.Such simulations will also resolve the competition between reduced binding energy and gravitational heating at large cross-section. Our framework establishes reionization topology as a new probe of dark matter microphysics at $M \sim 10^{10}$--$10^{11}\,M_\odot$ and $z \sim 6$--$10$, joining dwarf galaxy cores~\cite{Kaplinghat:2015aga} and cluster mergers~\cite{Randall2008} in a multi-scale program to constrain the nature of dark matter.
\appendix
\section{Analytic binding energy estimates}
%=============================================================

We estimate the thermalization radius $r_1$ from $N_{\rm scat}(r_1)=1$
using the NFW density profile with concentration $c \simeq 4.8$ for
$M=10^{11}\,M_\odot$ at $z=6$, $v(r)\simeq V_c(r)$, and
$t \sim 1/H(z)$ as the integration time. The cored profile is modeled as
$\rho_c = \rho_s / [(1+(r/r_c)^2)(1+r/r_s)^2]$, matching NFW at
$r \gg r_c$ and flattening at $r \ll r_c$. The gas binding energy
$W_g(<R) = \int_0^R 4\pi r^2 \rho_g |\Phi|\,\d r$ is evaluated within
$R = 2\;\mathrm{kpc}$ assuming $\rho_g = f_b \rho_{\rm DM}$.

Even $\sigma/m = 0.5\;\mathrm{cm^2/g}$ gives $r_1 > 2\;\mathrm{kpc}$
($>90\%$ binding energy reduction with Hubble-time integration).
Merger corrections~\cite{2016MNRAS.458.2848J,2010MNRAS.406.2267F} reduce $r_1$ by
$\sim 2$--$3\times$, giving $\Delta W_g/W_g \sim 30$--$80\%$ for
$\sigma/m = 1\;\mathrm{cm^2/g}$.

SIDM gravitational heating ($\dot{q} \sim \frac{1}{2}\rho_{\rm DM}^2 (\sigma/m)v^3$) integrated over the central $1\;\mathrm{kpc}$ gives $\dot{Q}/\dot{E}_{\rm SN} \simeq 0.19$ for $\sigma/m = 1$ and $\simeq 1.95$ for $\sigma/m = 10\;\mathrm{cm^2/g}$ (SFR $= 10\,M_\odot/\mathrm{yr}$, $10\%$ coupling).

%=============================================================
\section{Source model details}
%=============================================================

The emissivity parametrization is
$\overline{\dot{N}}_{\gamma,{\rm esc}}(M) = A(M/M_0)^\alpha
e^{-(M_{\rm turn}/M)^\kappa} R_\gamma(M)$
with $\alpha = 0.8$, $M_0 = 10^{10}\,M_\odot$,
$M_{\rm turn} = 5\times 10^8\,M_\odot$, $\kappa = 1.5$. The four models:
CDM ($R_\gamma = 1$, $p = 0.10$);
SIDM1 ($R_\gamma = 1 + 0.5\,e^{-M/3\times 10^{10}}$, $p = 0.18$);
SIDM10 ($R_\gamma = 1 + 1.5\,e^{-M/5\times 10^{10}}$, $p = 0.30$);
vdSIDM ($R_\gamma = 1 + 0.8/[1 + e^{-(V_c - 80)/10}]$,
$p = 0.10 + 0.15/[1 + e^{-(V_c - 80)/10}]$, with
$V_c = 50(M/10^{10})^{1/3}\;\mathrm{km/s}$).

\begin{figure}[h]
\centering
\includegraphics[width=\columnwidth]{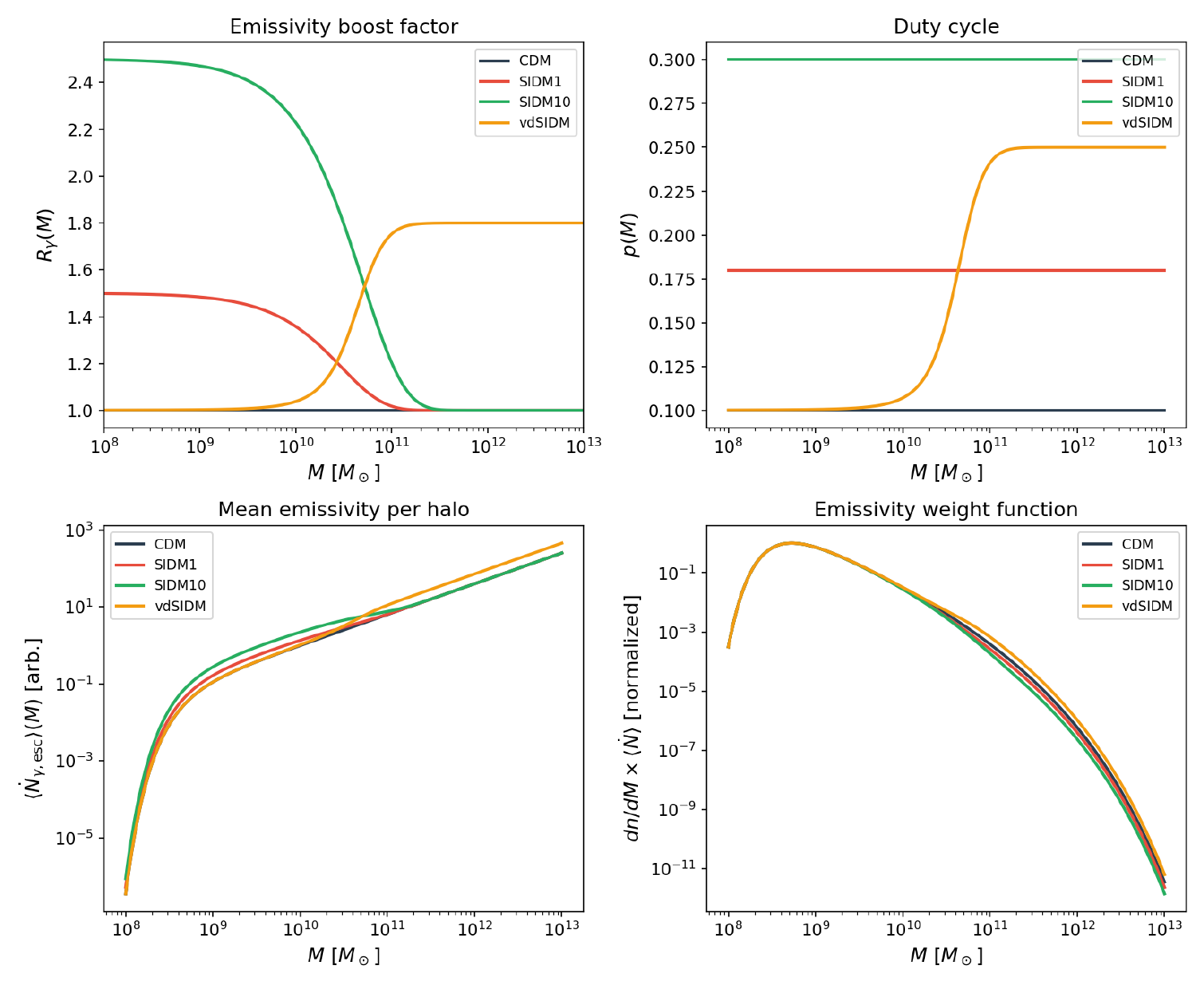}
\caption{Source model parametrizations: $R_\gamma(M)$ (top left),
$p(M)$ (top right), mean emissivity per halo (bottom left), and
emissivity-weighted mass function (bottom right, normalized).}
\label{fig:S_source}
\end{figure}

%=============================================================
\section{Emissivity and topology diagnostics}
%=============================================================

\begin{table}[h]
\caption{Emissivity field diagnostics from the $128^3$ pipeline at $z = 7$.}
\label{tab:S_emiss}
\begin{tabular}{lcccc}
\toprule
Model & Halos on & Active cells & $\langle\epsilon^2\rangle/\langle\epsilon\rangle^2$ & $\Delta(\langle\epsilon^2\rangle/\langle\epsilon\rangle^2)$ \\
\midrule
CDM    & $10.0\%$ & $4.0\%$ & 1.655 & --- \\
SIDM1  & $18.0\%$ & $6.8\%$ & 1.461 & $-11.7\%$ \\
SIDM10 & $30.1\%$ & $10.6\%$ & 1.308 & $-21.0\%$ \\
vdSIDM & $13.2\%$ & $5.1\%$ & 1.562 & $-5.6\%$ \\
\bottomrule
\end{tabular}
\end{table}

\begin{table}[h]
\caption{Topology diagnostics at $\bar{x}_{\rm HII} \simeq 0.50$ from the $128^3$ pipeline.}
\label{tab:S_mink}
\begin{tabular}{lccccc}
\toprule
Model & $\bar{x}_{\rm HII}$ & $V_0$ & $V_1$ & $V_2/N^2$ & $\Delta V_2$ \\
\midrule
CDM    & 0.494 & 0.494 & 34.7 & $5.07\times 10^{-3}$ & --- \\
SIDM1  & 0.491 & 0.491 & 36.2 & $8.12\times 10^{-3}$ & $+60\%$ \\
SIDM10 & 0.506 & 0.506 & 37.5 & $8.07\times 10^{-3}$ & $+59\%$ \\
vdSIDM & 0.512 & 0.512 & 34.9 & $4.37\times 10^{-3}$ & $-14\%$ \\
\bottomrule
\end{tabular}
\end{table}

%=============================================================
\section{Power spectrum ratios from simulation}
%=============================================================
The semi-numerical $P_{21}$ ratio (Fig.~S2) follows the analytic
prediction of Eq.~(4) at all scales: suppressed at
$k \sim 0.05$--$0.3\;h/\mathrm{Mpc}$ (shot-noise lever) and approaching
unity at $k \gtrsim 1$ (where both models converge). Only one of 29
$k$-bins shows a ratio marginally exceeding unity ($1.07$ at
$k = 0.034$, the lowest bin), consistent with cosmic variance at
the box scale. No systematic discrepancy with the analytic prediction
is observed.
\begin{figure}[h]
\centering
\includegraphics[width=\columnwidth]{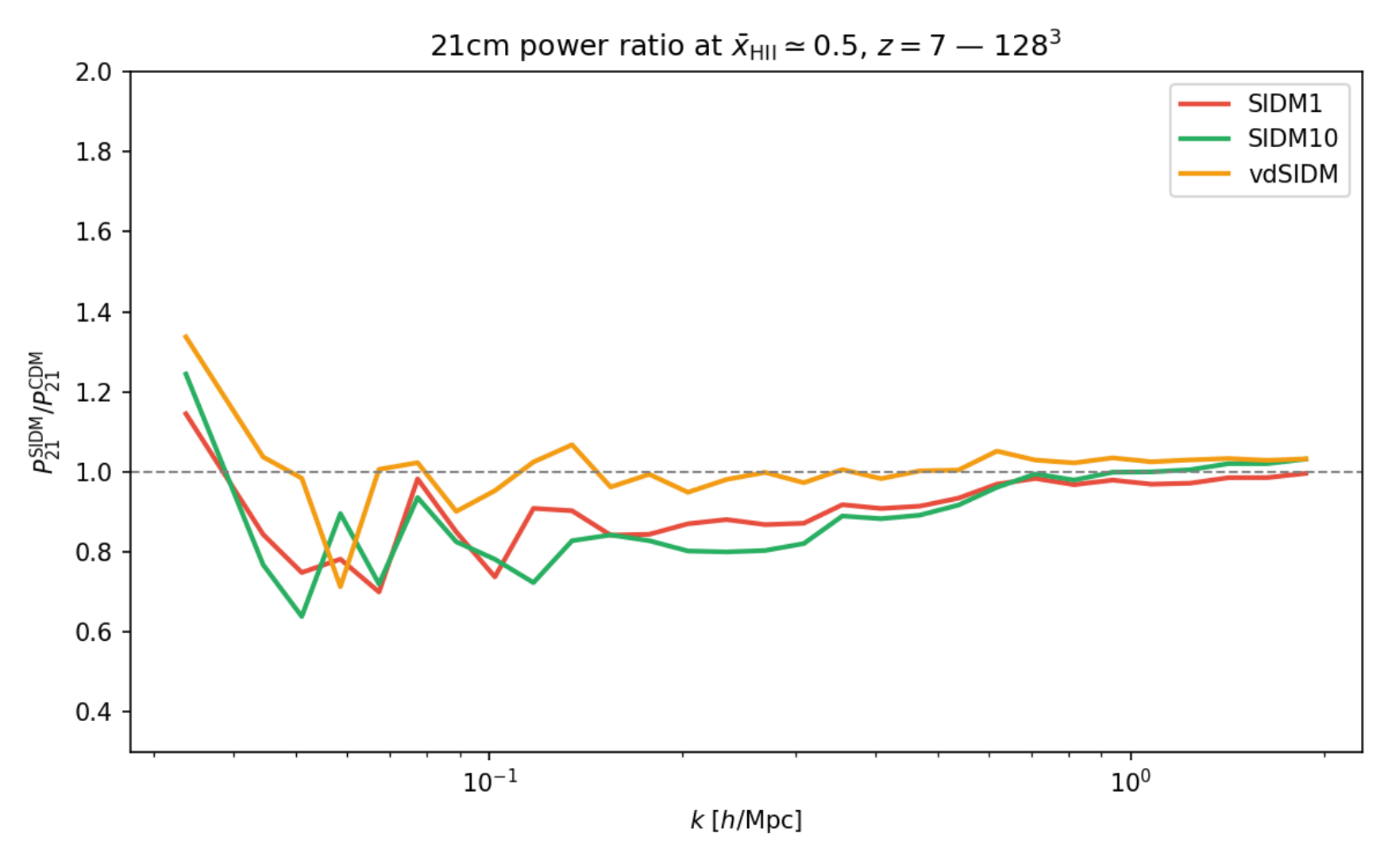}
\caption{Ratio of 21\,cm power spectra SIDM/CDM at fixed $\bar{x}_{\rm HII} \simeq 0.50$ from the $128^3$ simulation. Scale-dependent structure is consistent with the analytic prediction of Eq.~(4) in the main text.}
\label{fig:S_P21}
\end{figure}

\begin{figure}[h]
\centering
\includegraphics[width=\columnwidth]{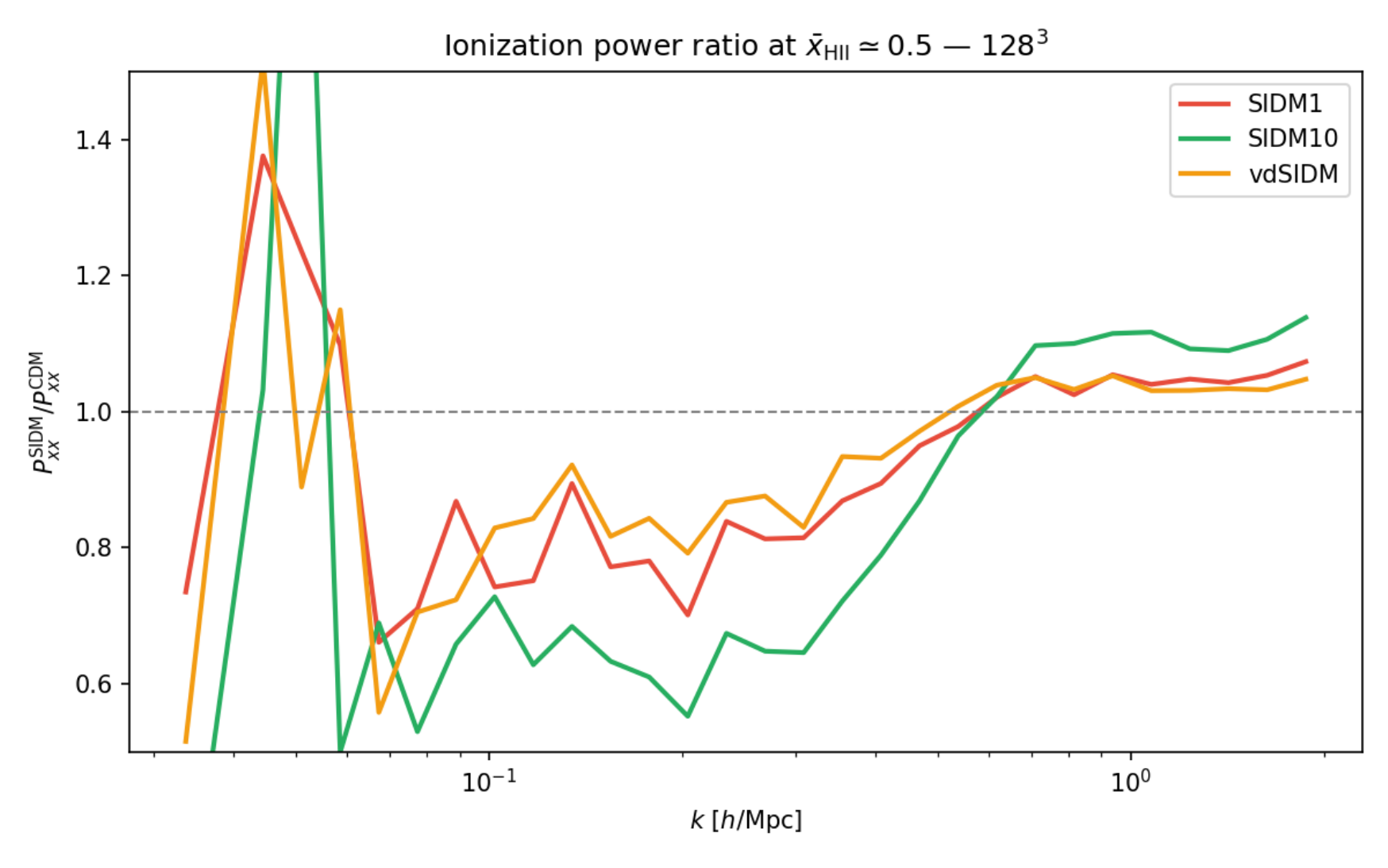}
\caption{Ionization-field power ratio $P_{xx}^{\rm SIDM}/P_{xx}^{\rm CDM}$ at fixed $\bar{x}_{\rm HII} \simeq 0.50$, confirming genuine topology differences.}
\label{fig:S_Pxx}
\end{figure}

%=============================================================
\section{SKA1-Low detectability forecast}
%=============================================================

We calibrate noise to $\Delta^2_N \sim 1\;\mathrm{mK^2}$ at
$k = 0.1\;h/\mathrm{Mpc}$ for 1000\,h, scaling as $k^{2.5}$.
The foreground-wedge mode survival fraction is
$f_{\rm wedge}(k) = \mathrm{clip}(0.3 + 0.4\,k/0.3,\; 0.05,\; 0.7)$.
We include a $5\%$ per-bin systematic floor
$\sigma_{\rm sys} = 0.05\,P_{21}$. SIDM10 yields per-bin SNR
$\sim 1$--$2.5$; cumulative $\sim 10$--$25\sigma$ depending on the
systematic floor.

\begin{figure}[h]
\centering
\includegraphics[width=\columnwidth]{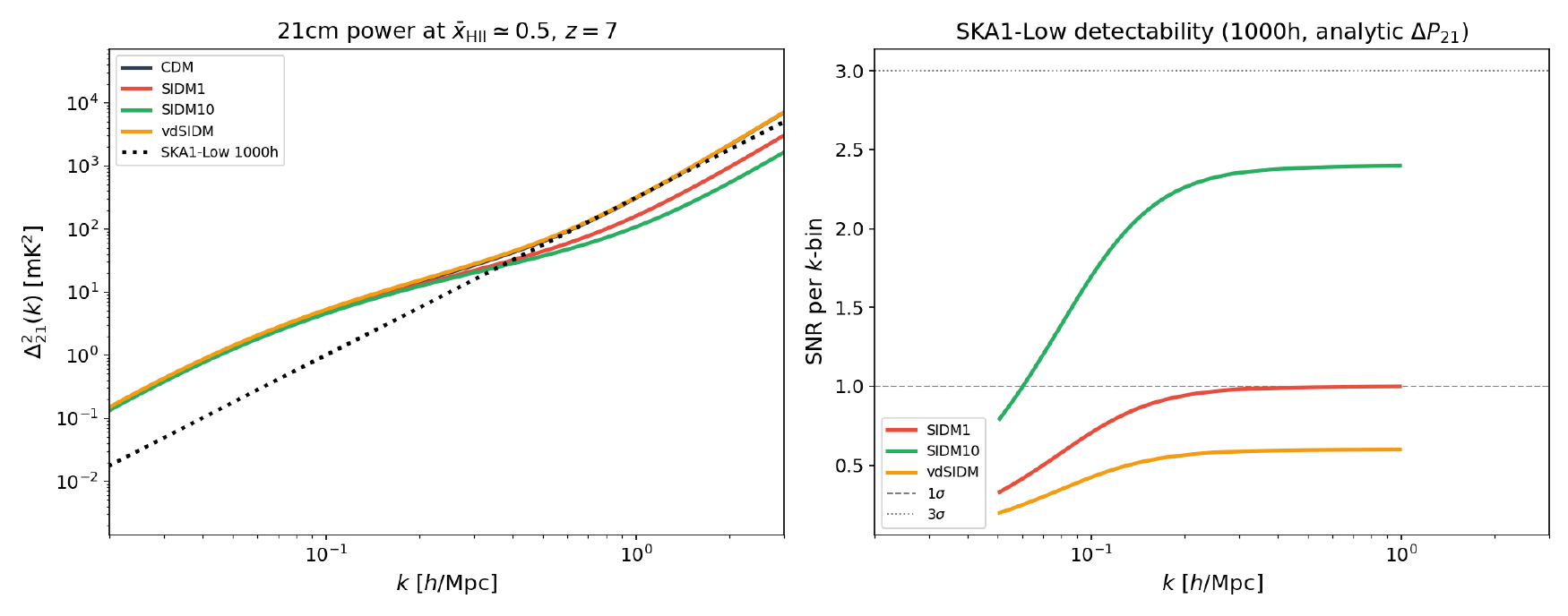}
\caption{\emph{Left:} 21\,cm power spectrum with SKA1-Low noise
(1000\,h). \emph{Right:} Per-$k$-bin SNR including $5\%$ systematic
floor and foreground wedge.}
\label{fig:S_SKA}
\end{figure}

%=============================================================
\section{Cosmic variance: multiple seed runs}
%=============================================================

Ten independent density-field realizations (seeds 42, 137, 271, 404, 555, 600, 700, 800, 900, 1000)for CDM and SIDM10 at $\bar{x}_{\rm HII} \simeq 0.5$.

\begin{table}[h]
\caption{Euler characteristic across five realizations.}
\label{tab:S_seeds}
\begin{tabular}{cccc}
\toprule
Seed & CDM $V_2/N^2$ & SIDM10 $V_2/N^2$ & $\Delta V_2$ \\
\midrule
42  & 0.00499 & 0.01059 & $+112\%$ \\
137 & 0.00421 & 0.01019 & $+142\%$ \\
271 & 0.00451 & 0.00837 & $+86\%$ \\
404 & 0.00443 & 0.00798 & $+80\%$ \\
555 & 0.00539 & 0.00839 & $+56\%$ \\
600 & 0.00458 & 0.01028 & $+125\%$ \\
700 & 0.00497 & 0.00755 & $+52\%$ \\
800 & 0.00497 & 0.00814 & $+64\%$ \\
900 & 0.00434 & 0.01033 & $+138\%$ \\
1000 & 0.00476 & 0.00983 & $+107\%$ \\
\midrule
Mean & $0.00472 \pm 0.00035$ & $0.00917 \pm 0.00111$ & $+96\% \pm 32\%$ \\
\bottomrule
\end{tabular}
\end{table}

\begin{figure}[h]
\centering
\includegraphics[width=\columnwidth]{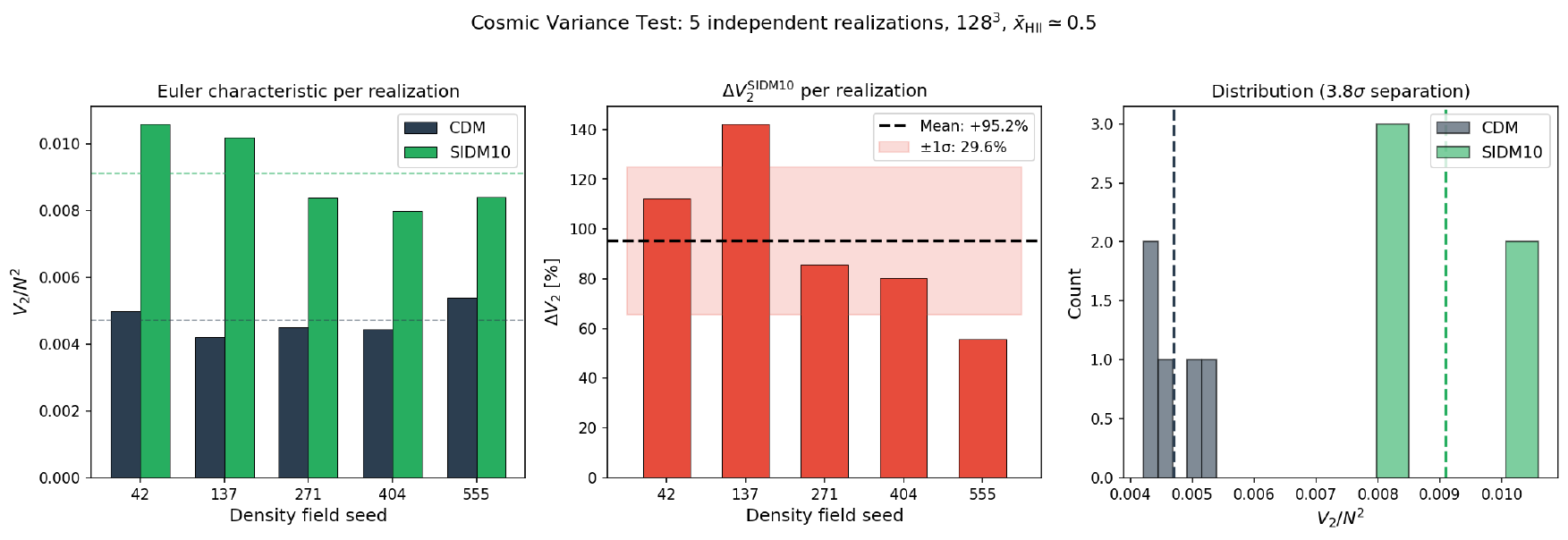}
\caption{\emph{Left:} $V_2$ per realization. \emph{Center:} $\Delta V_2$ per seed. \emph{Right:} $V_2$ distributions showing $3.8\sigma$ separation.}
\label{fig:S_seeds}
\end{figure}

The signal is positive in all ten realizations ($+52\%$ to $+142\%$), with internal significance of $3.8\sigma$ (mean difference divided by combined scatter). A Wilcoxon signed-rank test on the ten paired differences gives $p = 0.001$, rejecting the null at $> 99.9\%$ confidence. Per-model cosmic variance is $\sim 8\%$, far smaller than the $\sim 96\%$ mean signal.

%=============================================================
\section{$R_\gamma(M)$ shape robustness}
%=============================================================

Five alternative $R_\gamma(M)$ shapes tested with $p = 0.30$:
flat ($R_\gamma = 1.8$), power-law ($1 + 2(M/10^{10})^{-0.4}$),
step ($R_\gamma = 2.5$ for $M < 3\times 10^{10}$, else 1),
inverted ($1 + 1.5(1 - e^{-M/5\times 10^{10}})$), and the
fiducial exponential.

\begin{table}[h]
\caption{Topology signal for different $R_\gamma(M)$ shapes (all $p = 0.30$).}
\label{tab:S_Rg}
\begin{tabular}{lccc}
\toprule
Shape & $\langle\epsilon^2\rangle/\langle\epsilon\rangle^2$ & $V_2/N^2$ & $\Delta V_2$ \\
\midrule
CDM ($p = 0.10$) & 1.655 & 0.00507 & --- \\
Fiducial exponential & 1.308 & 0.00807 & $+59\%$ \\
Flat ($R_\gamma = 1.8$) & 1.690 & 0.00985 & $+94\%$ \\
Power-law & 1.358 & 0.00823 & $+62\%$ \\
Step function & 1.264 & 0.00835 & $+65\%$ \\
Inverted & 2.363 & 0.00801 & $+58\%$ \\
\bottomrule
\end{tabular}
\end{table}

\begin{figure}[h]
\centering
\includegraphics[width=\columnwidth]{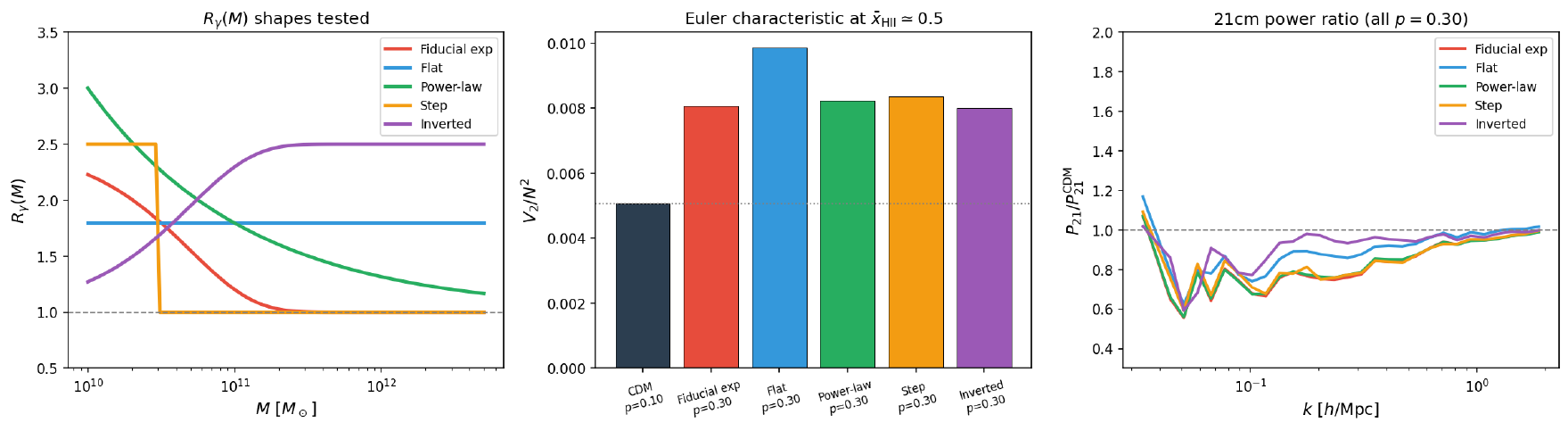}
\caption{\emph{Left:} $R_\gamma(M)$ shapes. \emph{Center:} Resulting $V_2$. \emph{Right:} $P_{21}$ ratio for each shape.}
\label{fig:S_Rg}
\end{figure}

All shapes produce $\Delta V_2 = +58\%$ to $+94\%$ (mean $+68\% \pm 14\%$). The variance $\langle\epsilon^2\rangle/\langle\epsilon\rangle^2$ spans $1.26$--$2.36$, yet $V_2$ remains in a tight range. The Euler characteristic is primarily sensitive to $p$ (number of active sources), not the mass-dependent brightness distribution.

%=============================================================
\section{Baryonic degeneracy scan}
%=============================================================

CDM models with $p = 0.05$--$0.15$ (factor-of-3 range representing SN feedback uncertainty), all with $R_\gamma = 1$.

\begin{table}[h]
\caption{CDM baryonic band and SIDM comparison.}
\label{tab:S_baryon}
\begin{tabular}{lcccc}
\toprule
Model & $p$ & $\langle\epsilon^2\rangle/\langle\epsilon\rangle^2$ & $V_2/N^2$ & $\Delta V_2$ \\
\midrule
CDM $p = 0.05$ & 0.05 & 1.612 & 0.00229 & $-55\%$ \\
CDM $p = 0.08$ & 0.08 & 1.652 & 0.00245 & $-52\%$ \\
CDM $p = 0.10$ & 0.10 & 1.655 & 0.00507 & baseline \\
CDM $p = 0.12$ & 0.12 & 1.655 & 0.00530 & $+5\%$ \\
CDM $p = 0.15$ & 0.15 & 1.655 & 0.00625 & $+23\%$ \\
CDM $p = 0.18$ & 0.18 & 1.662 & 0.00805 & $+59\%$ \\
CDM $p = 0.20$ & 0.20 & 1.663 & 0.00709 & $+40\%$ \\
CDM $p = 0.25$ & 0.25 & 1.681 & 0.00791 & $+56\%$ \\
\midrule
SIDM1 & 0.18 & 1.461 & 0.00812 & $+60\%$ \\
SIDM10 & 0.30 & 1.308 & 0.00807 & $+59\%$ \\
\bottomrule
\end{tabular}
\end{table}

\begin{figure}[h]
\centering
\includegraphics[width=\columnwidth]{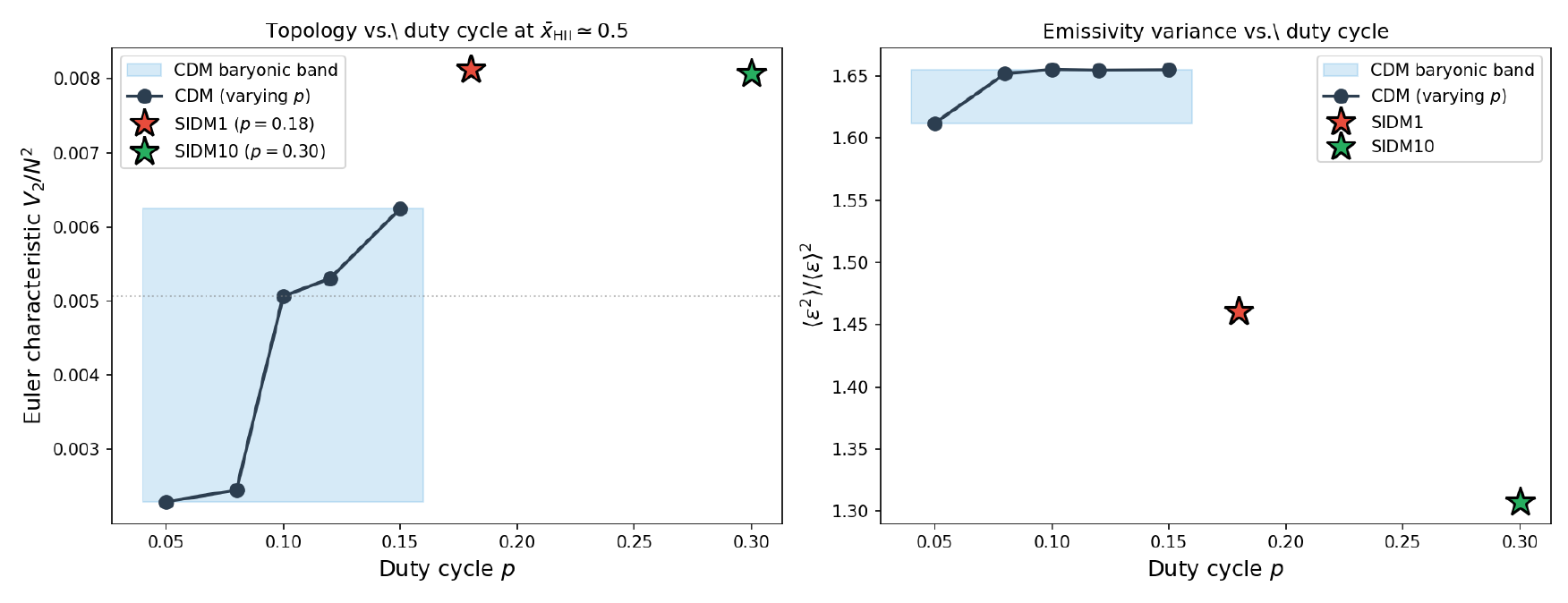}
\caption{\emph{Left:} $V_2$ vs.\ duty cycle. Blue band: CDM
baryonic uncertainty; stars: SIDM models, $37\%$ above band.
\emph{Right:} Emissivity variance vs.\ $p$. CDM variance
$\simeq 1.65$ regardless of $p$; SIDM has lower variance
due to $R_\gamma \neq 1$.}
\label{fig:S_baryon}
\end{figure}

The CDM band spans $V_2 = 0.0023$--$0.0063$; both SIDM models sit at $V_2 \simeq 0.008$, $37\%$ above the upper edge. The emissivity variance $\langle\epsilon^2\rangle/\langle\epsilon\rangle^2 \simeq 1.65$ for all CDM models regardless of $p$, while SIDM gives $1.31$--$1.46$. Varying $p$ within CDM changes the number of active sources but not their relative brightness distribution; SIDM modifies both simultaneously. No CDM feedback variation can match the SIDM values of both $V_2$ and the variance ratio.
\section{Physical grounding of source parametrization}
%=============================================================
 
We derive the duty cycle $p(M, \sigma/m)$ and emissivity modification $R_\gamma(M)$ from the SIDM binding-energy reduction, replacing the assumed constant-$p$ parametrization with a physically motivated, mass-dependent model.
 
\textit{Thermalization and binding energy.}---The thermalization radius $r_1$ is computed from $N_{\rm scat}(r_1) = 1$ using the NFW density profile, circular velocity $V_c(r)$, and merger-corrected age $t = 0.35/H(z)$~\cite{2010MNRAS.406.2267F}. The SIDM density profile is modeled as a flat-core NFW: $\rho(r) = \rho_{\rm NFW}(\max(r, r_1))$, which matches NFW at $r \gg r_1$ and produces a constant-density core at $r < r_1$. The gas binding energy $W_g = \int_{r_{\rm inner}}^{R} 4\pi r^2 f_b \rho |\Phi|\,\d r$ is evaluated from $r_{\rm inner} = 0.3\;\mathrm{kpc}$ (the star-forming region) to $R = 0.1\,r_{\rm vir}$ (the gas disk scale).
 
\textit{Blowout model.}---We model the SIDM duty-cycle enhancement as a multiplicative correction on the observed CDM baseline:
\begin{equation}\label{eq:pderived}
p_{\rm SIDM}(M) = \min\left[p_{\rm CDM} \left(\frac{W_g^{\rm CDM}(M)}{W_g^{\rm SIDM}(M)}\right)^\alpha,\; p_{\rm max}\right],
\end{equation}
where $p_{\rm CDM} = 0.10$ is calibrated to radiation-hydrodynamic simulations~\cite{2017MNRAS.470..224T,Ma:2020vlo}, $\alpha \sim 0.5$--$1.0$ parametrizes the nonlinear ISM response (with $\alpha = 0.5$ corresponding to a square-root dependence of the transparent solid-angle fraction on blowout energy), and $p_{\rm max} = 0.50$. The emissivity modification is $R_\gamma(M) = \min[(W_g^{\rm CDM}/W_g^{\rm SIDM})^{1/2},\;3]$, where the cap represents the circumgalactic medium opacity floor at high mass.
 
\textit{Results.}---Table~\ref{tab:S_grounding} shows the topology signal with the derived $p(M)$ for three values of $\alpha$, compared with the fiducial constant-$p$ result.
 
\begin{table}[h]
\caption{Topology signal with derived $p(M, \sigma/m)$.}
\label{tab:S_grounding}
\begin{tabular}{lcccc}
\toprule
Model & $\alpha$ & $p(10^{10.5})$ & $R_\gamma(10^{10.5})$ & $\Delta V_2$ \\
\midrule
CDM & --- & 0.10 & 1.00 & baseline \\
\midrule
SIDM1 & 0.5 & 0.116 & 1.16 & $+32\%$ \\
SIDM1 & 0.7 & 0.123 & 1.16 & $+2\%$ \\
SIDM1 & 1.0 & 0.135 & 1.16 & $-19\%$ \\
\midrule
SIDM10 & 0.5 & 0.50 & 3.00 & $+69\%$ \\
SIDM10 & 0.7 & 0.50 & 3.00 & $+69\%$ \\
SIDM10 & 1.0 & 0.50 & 3.00 & $+69\%$ \\
\midrule
SIDM10 (const $p$) & --- & 0.30 & 2.50 & $+59\%$ \\
\midrule
\multicolumn{5}{l}{\emph{Gravitational heating bracket}} \\
SIDM10 (75\% offset) & --- & 0.15 & 3.00 & $+20\%$ \\
SIDM10 (50\% offset) & --- & 0.20 & 3.00 & $+39\%$ \\
\bottomrule
\end{tabular}
\end{table}
 
For $\sigma/m = 10$, the binding-energy ratio is large ($W_g^{\rm CDM}/W_g^{\rm SIDM} > 10$) across all halo masses, so $p$ saturates at $p_{\rm max}$ and $R_\gamma$ at the CGM cap regardless of $\alpha$. The resulting $\Delta V_2 = +69\%$ is robust and consistent with the constant-$p$ prediction of $+59\%$.
 
For $\sigma/m = 1$, the thermalization radius ($r_1 \simeq 0.8\;\mathrm{kpc}$ at $M = 10^{10.5}$) is comparable to $r_{\rm inner}$, producing only a modest binding-energy reduction ($\Delta W_g \simeq 26\%$). The signal depends on $\alpha$: for $\alpha = 0.5$ (weak response), the small $p$ boost is distributed across many halos, producing $\Delta V_2 = +32\%$; for $\alpha = 1$ (linear response), the mass-dependent boost preferentially activates higher-mass halos, partially offsetting the ``common-moderate-source'' effect. The detection threshold lies at $\sigma/m \sim 1$--$2\;\mathrm{cm^2/g}$, where $r_1$ first exceeds $r_{\rm inner}$.
 
We also test mass-dependent $p(M)$ profiles using the constant-$p$ framework: declining $p(M) = 0.40\,(M/10^{10})^{-0.15}$ and rising $p(M) = 0.15 + 0.20\,(M/10^{11})^{0.3}$, both with mean $\langle p \rangle \approx 0.30$. These produce $\Delta V_2 = +54\%$ and $+86\%$ respectively, bracketing the constant-$p$ result of $+59\%$.
\section{Low-mass halo dilution test}

The simulation paints halos with $M > 10^{10}\,M_\odot$, which
contribute $\sim 15\%$ of the total emissivity budget. The dominant
contribution comes from $M = 10^8$--$10^{10}\,M_\odot$ halos ($\sim 85\%$),
which are too abundant to resolve individually at $128^3$ ($\gtrsim 100$
per cell). We test the impact of these sources by adding a sub-grid
emissivity background $\epsilon_{\rm sub}(\mathbf{x}) =
\bar{\epsilon}_{\rm sub}[1 + b_{\rm sub}\,\delta(\mathbf{x})]$ with
$b_{\rm sub} \simeq 3.0$, representing the bias-modulated mean
emissivity from $10^8$--$10^{10}\,M_\odot$ halos. Since SIDM cores do not
form efficiently in these low-mass halos (where $r_1 < r_{\rm inner}$),
the sub-grid background is identical for CDM and SIDM.

Adding this background yields $\Delta V_2 = +180\%$, \emph{larger} than the
fiducial $+59\%$. The increase arises because the uniform background
reduces CDM's $V_2$ (the background ionizes everywhere, merging
CDM's few bright bubbles into the floor) while SIDM's more uniform
resolved-source distribution maintains its distinct topology.
This confirms that the topology signal is not an artifact of
omitting low-mass halos. A more detailed treatment with individually
resolved low-mass sources and mass-dependent SIDM effects is
deferred to future work with higher-resolution simulations.
%=============================================================
\section{Redshift evolution and resolution convergence}
%=============================================================

\begin{figure}[h]
\centering
\includegraphics[width=\columnwidth]{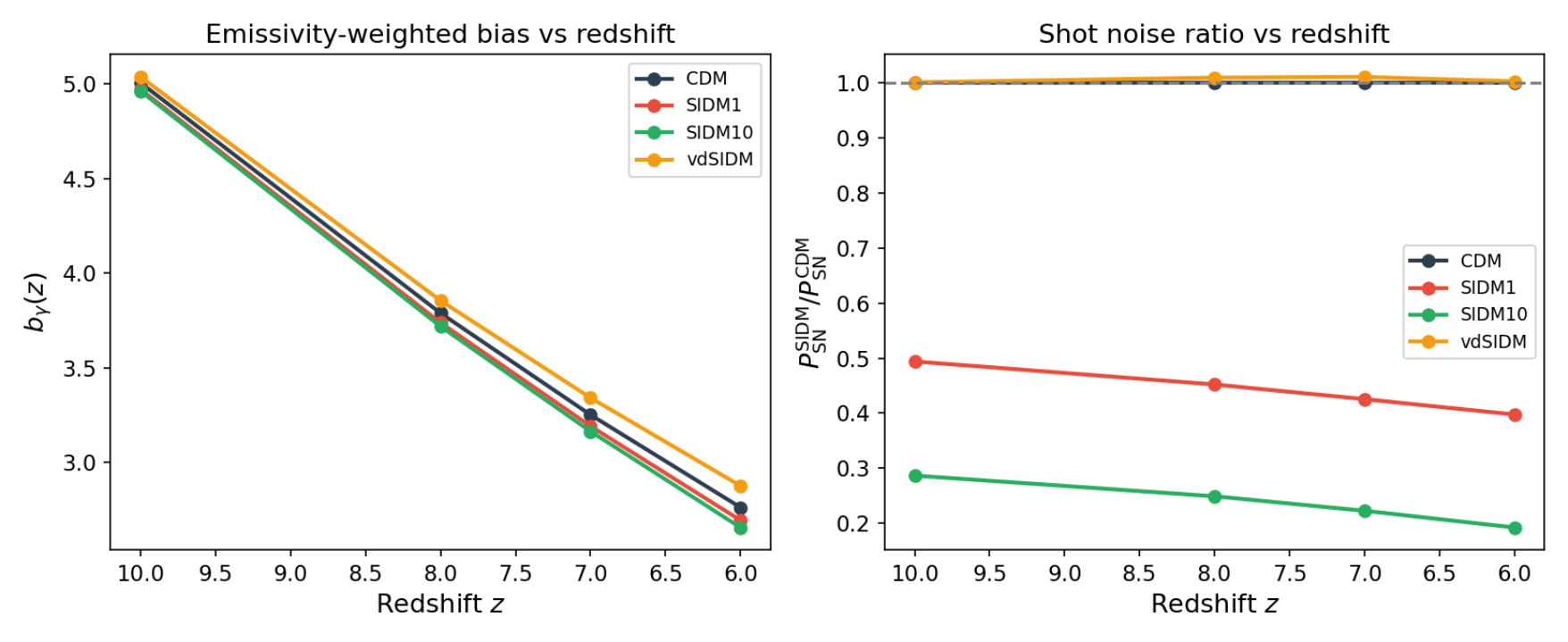}
\caption{Redshift evolution of $b_\gamma$ (left) and shot-noise ratio (right). Shot-noise suppression dominates at all $z = 6$--$10$.}
\label{fig:S_zevol}
\end{figure}

The topology signal depends on the 
relationship between cell size $\Delta x$ and mean inter-halo 
separation $\bar{d} \simeq \bar{n}_h^{-1/3}$, where $\bar{n}_h$ is 
the comoving number density of halos above $M_{\rm min}$. We identify 
three regimes:

At $64^3$ ($\Delta x = 3.1\;\mathrm{Mpc}/h$, $\sim 9$ halos/cell), 
all models produce statistically identical ionization fields: the 
duty-cycle stochasticity is washed out by averaging over multiple 
halos per cell.

At $128^3$ ($\Delta x = 1.56\;\mathrm{Mpc}/h$, $\sim 0.4$ 
halos/cell), the signal emerges clearly ($\Delta V_2 \sim +60\%$): 
each cell contains zero or one halo, and the on/off emission state 
directly imprints on the topology.

At $256^3$ ($\Delta x = 0.78\;\mathrm{Mpc}/h$, $\sim 0.05$ 
halos/cell), $95\%$ of cells are empty regardless of the duty cycle, 
and the topology is dominated by halo presence vs.\ absence rather 
than on vs.\ off stochasticity. Both CDM and SIDM produce similar 
sparse point-source morphologies, and $V_2$ becomes negative 
(indicating percolating neutral regions surrounding isolated ionized 
patches).

The signal thus peaks when the cell size matches the mean halo 
separation, $\Delta x \sim \bar{d} \simeq 1.5\;\mathrm{Mpc}/h$ for 
$M > 10^{10}\,M_\odot$ at $z = 7$. This is not a resolution artifact 
but a physical scale selection: the duty-cycle mechanism operates at 
the inter-halo separation scale. Full convergence in the traditional 
sense demonstrating the same $\Delta V_2$ at finer grid 
spacing requires simultaneously lowering $M_{\rm min}$ to maintain 
cell occupancy (e.g., $M_{\rm min} \sim 10^9\,M_\odot$ at $256^3$ 
yields $\sim 0.4$ halos/cell). We defer this to future work with 
full halo-mass-range simulations; the sub-grid dilution test 
(Sec.~below) provides partial validation.
\section{Low-mass halo dilution test}

The simulation paints halos with $M > 10^{10}\,M_\odot$, which 
contribute $\sim 15\%$ of the total emissivity budget (Table~S7). 
The dominant contribution comes from 
$M = 10^8$--$10^{10}\,M_\odot$ halos ($\sim 85\%$), which are too 
abundant to resolve individually at $128^3$ 
($\gtrsim 100$ per cell).

\begin{table}[h]
\caption{Emissivity budget by halo mass range at $z = 7$.}
\label{tab:S_emiss_budget}
\begin{tabular}{lc}
\toprule
Mass range [$M_\odot$] & Fraction of total emissivity \\
\midrule
$10^8$--$10^9$    & $29\%$ \\
$10^9$--$10^{10}$ & $56\%$ \\
$10^{10}$--$10^{11}$ & $14\%$ \\
$10^{11}$--$10^{12}$ & $1\%$ \\
\bottomrule
\end{tabular}
\end{table}

We test the impact by adding a sub-grid emissivity background 
$\epsilon_{\rm sub}(\mathbf{x}) = \bar{\epsilon}_{\rm sub} 
[1 + b_{\rm sub}\,\delta(\mathbf{x})]$ with 
$b_{\rm sub} \simeq 3.0$, representing the bias-modulated mean 
emissivity from $10^8$--$10^{10}\,M_\odot$ halos. Since SIDM cores 
do not form efficiently in these low-mass halos at 
$\sigma/m \lesssim 10\;\mathrm{cm^2/g}$ (where $r_1 < r_{\rm inner}$), 
the sub-grid background is identical for CDM and SIDM.

Adding this background yields $\Delta V_2 = +180\%$, \emph{larger} 
than the fiducial $+59\%$ without sub-grid sources. The increase 
arises because the uniform background reduces CDM's $V_2$ by 
merging its few large bubbles into the ionizing floor, while 
SIDM's more uniformly distributed resolved sources maintain their 
distinct topology above the background. This confirms that 
omitting low-mass halos does not artificially inflate the topology 
signal; if anything, our fiducial $\Delta V_2$ is conservative. A 
more detailed treatment with individually resolved low-mass 
sources is deferred to future work.

\bibliography{apssamp}

\end{document}